\documentclass[
  epj,
]{svjour}
\usepackage{graphicx, xcolor}
\usepackage{amsmath,bm,amsfonts}
\usepackage{braket}
\usepackage{cite}
\usepackage[
  bookmarks=true,
  colorlinks,
  linkcolor=blue,
  urlcolor=blue,
  citecolor=blue,
  plainpages=false,
  pdfpagelabels,
  final,
  breaklinks=true
]{hyperref}

\newcommand{\orcid}[1]{%
  ~\href{https://orcid.org/#1}{\includegraphics[width=8pt]{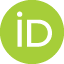}}%
  }

 \makeatletter
 \def\makeheadbox{{%
 }}
 \makeatother

\newcommand{\figref}[1]{\figurename~\ref{#1}}
\newcommand{\secref}[1]{Sec.~\ref{#1}}
\renewcommand{\eqref}[1]{Eq.~(\ref{#1})}

\newcommand{\ip}{\mathrm{I}_\mathrm{p}}
\newcommand{\up}{\mathrm{U}_\mathrm{p}}
\newcommand{\pb}{\mathbf{p}}

\newcommand{\rb}{\mathbf{r}}

\newcommand{\Ab}{\mathbf{A}}

\newcommand{\mel}[3]{\braket{#1|#2|#3}}

\begin{document}
\title{Conservation laws for Electron Vortices in Strong-Field Ionisation}
\author{Yuxin Kang\inst{1} \and 
	Emilio Pisanty\orcid{0000-0003-0598-8524}\inst{2,3} \and
	Marcelo Ciappina\orcid{0000-0002-1123-6460}\inst{3,4,5} \and
	Maciej Lewenstein\orcid{0000-0002-0210-7800}\inst{3,6} \and
	Carla Figueira de Morisson Faria\orcid{0000-0001-8397-4529}\inst{1} \and
	Andrew S Maxwell\orcid{0000-0002-6503-4661}\inst{1,3}
}
	
%
%
\institute{Department of Physics \& Astronomy, University College London, Gower Street, London WC1E 6BT, United Kingdom
	\and
Max Born Institute for Nonlinear Optics and Short Pulse Spectroscopy, Max-Born-Stra{\ss}e 2A, Berlin 12489, Germany
\and
ICFO - Institut de Ciencies Fotoniques, The Barcelona Institute of Science and Technology, Av. Carl Friedrich Gauss 3, 08860 Castelldefels (Barcelona), Spain
\and
Physics Program, Guangdong Technion -- Israel Institute of Technology, Shantou, Guangdong 515063, China
\and
Technion -- Israel Institute of Technology, Haifa, 32000, Israel
\and
ICREA, Pg. Lluís Companys 23, 08010 Barcelona, Spain
}
\date{Received: date / Revised version: date}
%
\abstract{
We investigate twisted electrons with a well defined orbital angular momentum, which have been ionised via a strong laser field. By formulating a new variant of the well-known strong field approximation, we are able to derive conservation laws for the angular momenta of twisted electrons in the cases of linear and circularly polarised fields. In the case of linear fields, we demonstrate that the orbital angular momentum of the twisted electron is determined by the magnetic quantum number of the initial bound state. The condition for the circular field can be related to the famous ATI peaks, and provides a new interpretation for this fundamental feature of photoelectron spectra. We find the length of the circular pulse to be a vital factor in this selection rule and, employing an effective frequency, we show that the photoelectron OAM emission spectra are sensitive to the parity of the number of laser cycles. This work provides the basic theoretical framework with which to understand the OAM of a photoelectron undergoing strong field ionisation.
} 
\maketitle
\section{Introduction}
\label{intro}

Since being first recognised in the classical context of tides \cite{whewell1833,whewell1836,berry2001} vortex phenomena have held an iconic status across a diverse range of disciplines \cite{hall1951,swithebank1969,choe2004,Sumino2012}. Of particular interest are the quantum mechanical versions of vortices, which can be found in a wide range of systems \cite{Allen1992,Blatter1994,Bliokh2007,Fetter2009,pitaevskiistringari2016book}. Current intense interest in phase vortices follows the first experimental observations of this phenomenon for photons \cite{Allen1992} and unbound electrons \cite{Bliokh2007} (for reviews on vortices in electrons see \cite{mcmorran2016,bliokh2017,lloyd2017}).
The unique topological properties of vortex states \cite{lloyd2017} render them fundamental in the study of structured wave fields, and have led to the inception of whole research areas, such as singular optics \cite{torres2011,andrews2012,gbur2016,rubinsztein2017}. For instance, a vortex state cannot transform to another by simple deformation such as stretching and compressing, or the addition of noise.  Furthermore, at its centre along the propagation axis, a vortex beam has a zero amplitude and an ill-defined phase. Currents around singularities imply that such states carry intrinsic orbital angular momenta (OAM), and thus can be used to influence the dynamical properties of physical systems \cite{garces2002,padgett2011}. 

The study of vortices in attosecond physics has great potential in controlling light and matter. In high-harmonic generation (HHG) it has been shown that optical vortices in the IR driving fields lead to optical vortices in the resulting UV light produced \cite{Zurch2012,hernandez-garcia2013,gariepy2014,hernandez-garcia2015,geneaux2016,rego2016,turpin2017,kong2017,gauthier2017,hernandez-garcia2017,Paufler2018,dorney2019,Pisanty2019,Pisanty2019conservation,rego2019}.
This has been exploited to allow a high degree of control over the light. In one example, UV light has been produced exhibiting torus knot topology \cite{Pisanty2019,Pisanty2019conservation} by sophisticated trefoil IR pulses, for which the orientation of the trefoil varies with the azimuthal angle. In another study~ \cite{rego2019}, extreme UV (EUV) light was imparted with time-varying OAM, leading to self-torque by employing two time-delayed IR pulses with different OAM. This was the first demonstration of self-torque in light \cite{barati2020} and could aid in probing systems with naturally time-varying OAM.

Work understanding the OAM of photoelectrons emitted in attosecond processes is still in its infancy. Initial studies include the exploration of high OAM values for quasi-relativistic field intensities \cite{Velez2018} and terahertz fields \cite{Velez2020} as well as using the OAM in rescattering to probe bound state structures \cite{Tolstikhin2019}. 
The OAM has also been studied indirectly via the coherent combination of pairs of vortex states that lead to the interference vortices \cite{NgokoDjiokap2015,NgokoDjiokap2016,yuan2016,NgokoDjiokap2018,Pengel2017,Pengel2017a,Li2018, Kerbstadt2019,Kerbstadt2019a, Armstrong2019b,Maxwell2020Faraday}, which occur for two counter rotating circularly polarised fields separated by a time delay.
However, ideas as advanced as knots or self-torque in electron vortices have not been explored. 
This can partly be attributed to the difficulty in experimental implementation; for example, no measurement scheme has been devised to detect the OAM of photoelectrons emitted in strong-field experiments.
Initial ideas on how to achieve this have been suggested in our recent publication \cite{Maxwell2020Faraday}, using both interferometric schemes and adapting existing methods \cite{Grillo2017} used in electron beams.
Such experimental development needs sufficient theoretical backing to guide the implementation and justify the cost, which at present is lacking. In this study we introduce a theoretical framework to derive analytical conservation laws and understand the basic dynamics of the OAM during strong field ionisation. Note in all cases we consider the OAM of the electron but \textbf{not} that of the laser fields.  The OAM we refer to in this work is associated with the OAM carried by free particles, described by Bessel functions and is the eigenvalue of $\hat{L}_z$, that is, the quantum magnetic number. The outgoing photo\allowbreak{}electron wavepacket in strong-field ionisation does not naturally form a beam, so the $z$-direction can be arbitrarily chosen. Throughout this work we use the convention chosen in \figref{fig:OAMAxis}.

We utilise the strong field approximation (SFA) \cite{amini2019}---often considered the workhorse of strong field physics---to describe the basic ionisation dynamics. The SFA in the form used here \cite{Becker2002Review,Faria2002,Maxwell2020Faraday} is an approximation, which primarily neglects the effects of the Coulomb potential in the continuum. The use of the SFA, over a more accurate model, is justified by the possibility of analytical solutions and the ease of interpreting the results. Thus, the SFA allows the basic laws of the OAM to be derived for strong field ionisation. Furthermore, we focus mostly on circularly polarised light, where the Coulomb effects are less significant, than those observed in linear polarisation \cite{faria_it_2020}. In a recent publication \cite{Maxwell2020Faraday}, we numerically computed OAM distributions using the SFA, QProp \cite{qprop2,qprop3} and the $R$-matrix with time dependence method \cite{moore2011,clarke2018,rmtcpc} for circularly polarised fields and confirmed that the SFA was able to qualitatively reproduce the key features found in these two more accurate numerical methods. This work differed from the present in that it focused on the interference effects due to employing two time-delayed counter-rotating circularly polarised IR fields. Furthermore, the OAM computations were performed numerically without full derivation of the analytical expressions and the SFA calculations employed a monochromatic field, whereas in this work we extend the model to include a $\sin^2$ envelope. 

The paper is structured as follows. In \secref{sec:key_results} key theoretical results of the SFA (\secref{sec:key_results:SFA}) and vortex states (\secref{sec:key_results:vortex}) are given.  Next, in \secref{sec:matrix_element} we incorporate the OAM into the SFA, starting with expressions for a general field (\secref{sec:matrix_element:general}). Following this, we derive analytical conditions for the cases of a linear field (\secref{sec:matrix_element:linear}) and a monochromatic circular field; in the latter we do this both without (\secref{sec:matrix_element:circular}) and with (\secref{sec:matrix_element:circularSPA}) the use of the saddle point approximation.
In \secref{sec:experiment_detection} numerical results are presented for a circular $\sin^2$ laser field. Therein we derive an effective frequency to extend the analytic condition for a monochromatic field to work for the $\sin^2$ case.
Finally, in \secref{sec:conclusions} we present our conclusions.

\section{Key results from the strong field approximation and vortex states}
\label{sec:key_results}
\sloppy
In this section we provide some key results from the SFA and for electron vortex states, necessary for understanding the OAM derivation. Throughout the article we use atomic units unless otherwise stated. 
Both cylindrical $(\hat{e}_{\parallel},\hat{e}_{\perp},\hat{e}_{\phi} )$ and Cartesian $(\hat{e}_{x},\hat{e}_{y},\hat{e}_{z} )$ coordinates will be used in this article, they will always be aligned such that $\hat{e}_{\parallel}=e_{z}$ and any radial quantity will be given by e.g. $r_{\perp}=\sqrt{r_x^2+r_y^2}$. We will consider specific cases where the laser field polarisation is parallel to $\hat{e}_{\parallel}$ [linearly polarised] and in the $xy$-plane [circularly polarised].

\subsection{Key results from SFA }
\label{sec:key_results:SFA}
Within the SFA S-matrix formalism \cite{Becker2002Review,becker1997} the transition amplitude for direct strong field ionisation from the bound state $\ket{\psi_0(t')}$ to a final continuum state $\ket{\psi_{f}(t)}$ is given by the following expression \cite{Becker2002Review,becker1997}

\begin{equation}{\label{eq:transition-ampl-gen}}
M_{f}=-i\lim\limits_{t\rightarrow\infty}\int_{-\infty}^{t} dt'\mel{\psi_{f}(t)}{U_{v}(t,t')V}{\psi_{0}(t')},
\end{equation}
where the time evolution is approximated by the Volkov operator
\begin{equation}\label{eq:VolkovVortexState}
U_{v}(t,t')=\int d^{3}\pb e^{\frac{-i}{2}\int_{t'}^{t}(\pb+\Ab(\tau)^{2})d\tau}\ket{\pb+\Ab(t)}\bra{\pb+\Ab(t')}.
\end{equation} 	
Combining Eqs.
(\ref{eq:VolkovVortexState}) and (\ref{eq:transition-ampl-gen}) and taking $\Ab(t)=0$ we get the following for the transition amplitude
\begin{align}\label{eq:FullyExpandedTransitionAmplitude}
M_{f}&=\lim\limits_{t\rightarrow\infty}
\int d^{3}\pb'\;		
e^{-iS(\pb,t)}\braket{\psi_{f}|\pb'}M(\pb)
\intertext{with}
M(\pb)&=
-i\int_{-\infty}^{\infty}dt'e^{iS(\pb,t')}d(\pb,t'),
\label{eq:transition-ampl-plane}
\intertext{where}
d(\pb,t')&=\braket{\pb+\Ab(t')|V|\psi_{0}}
\end{align} 
and $S(\pb,t)$ and $S(\pb,t')$ are the upper and lower limit of the semi-classical action, respectively, both given by 
\begin{align}
S(\pb,t)=\ip t +\frac{1}{2}\int^{t}_{-\infty} d \tau (\pb+\Ab(\tau))^2.
\end{align}	
A plane wave momentum state is commonly used for the final continuum state $\ket{\psi_{f}}\rightarrow\ket{\psi_{\pb}}$, which leads to $\braket{\psi_f|\pb'}\rightarrow\braket{\pb|\pb'}=\delta(\pb'-\pb)$ and therefore $M_f\rightarrow M(\pb)$, where \eqref{eq:transition-ampl-plane} gives the definition of $M(\pb)$.	
However, in this work, a Bessel beam vortex state will be used as final continuum state instead of a plane wave. This will enable the development of an analytical model for the OAM of the outgoing photoelectron.   
In order to proceed we will introduce some properties of the Bessel beam.

\subsection{Key results from electron vortex state}
\label{sec:key_results:vortex}
Vortex states are topologically distinct both from plane waves and vortex states with a different orbital angular momentum $l$ \cite{lloyd2017}; this means two vortex states cannot be transformed into one another via continuous deformations.  At its centre along the propagation axis, the vortex has a zero amplitude and undefined phase. In order to enforce the  single-valued wave function, a quantised phase factor $e^{il\phi}$ is needed, where $\phi$ is the azimuthal angle and $l$ is a topological charge with integer value known as the orbital angular momentum (OAM). The general form of the Bessel beam electron-vortex is \cite{bliokh2017}
\begin{equation}\label{eq:BesselSpace}
\braket{\rb|\psi_{l}(t)}=N_{l}J_{l}(p_{\perp}r_{\perp})e^{il\phi}e^{i p_{\parallel}r_{\parallel}}e^{-i\omega t}.
\end{equation}
Here, $N_{l}$ is a normalisation factor and $J_l(p_{\perp}r_{\perp})$ is the Bessel function of the first kind. Ignoring the time dependence, the Fourier transform is 
\begin{equation}\label{eq:BesselMomentum}
\braket{\pb'|\psi_{l}}=\frac{i^{-l}e^{i l\phi'}}{2\pi p_{\perp}}\delta(p'_{\parallel}-p_{\parallel}) \delta(p'_{\perp}-p_{\perp}),
\end{equation}
where $(p_{\parallel},p_{\perp},\phi)$ are the cylindrical coordinates of $\pb$.
Inserting this into \eqref{eq:FullyExpandedTransitionAmplitude} gives
\begin{align}
&M_{l}(p_{\parallel},p_{\perp},t)
=\notag \\
&\frac{i^{l}}{2\pi}\int_{-\pi}^{\pi} d\phi'e^{-i S(p_{\parallel},p_{\perp},\phi',t)} e^{-il\phi'}
M(p_{\parallel},p_{\perp},\phi'),
\label{eq:VortexTransition1}
\end{align}
where $M(p_{\parallel},p_{\perp},\phi')$ is the transition amplitude \eqref{eq:transition-ampl-plane} written in cylindrical coordinates,
which have been used as they are natural for vortex states. Now the vortex state is incorporated into the SFA framework. Next, we will compute more explicit expressions for the transition amplitude.

\section{Matrix Element Calculations }
\label{sec:matrix_element}
In this section, the transition matrix element will be derived for a general laser field. We will examine the specific cases of a linear field and a circular monochromatic field, as well as the saddle point approximation.

\subsection{Coordinate Systems}
\begin{figure}
	\includegraphics[width=\columnwidth]{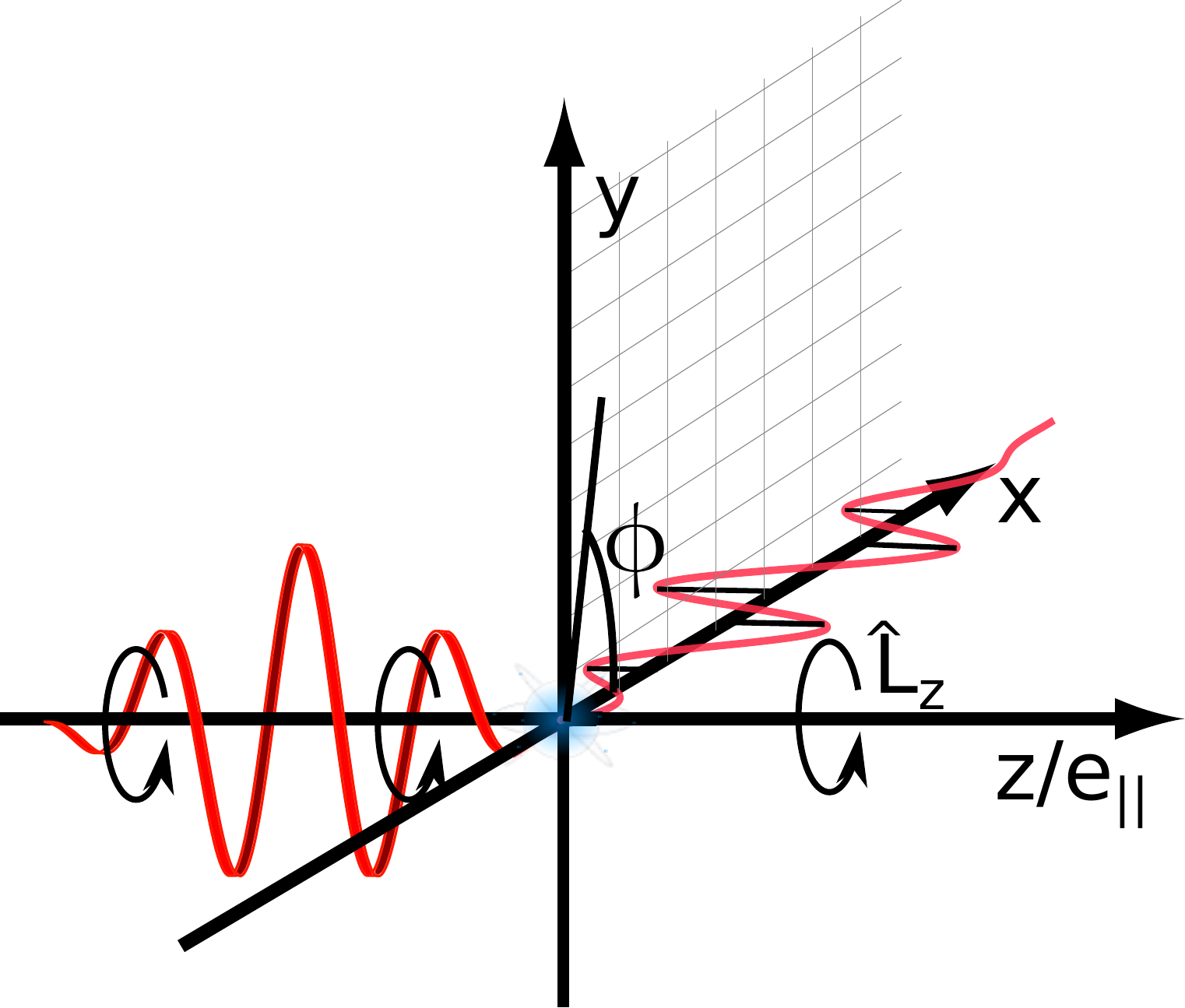}
	\caption{Depiction of the two laser fields and the coordinate system use throughout this article. The circularly polarised field propagates along the z-axis and its electric field vector rotates in the xy-plane (marked by a square grid). In contrast, the linearly polarised field propagates along the x-axis and its electric field vector oscillates along the z/ $\hat{e}_{||}$ direction. The angle $\phi$ is defined to be in the xy-plane (i.e. the angle about the z-axis) and starts from the x-axis. The orbital angular momentum is defined to be around the z-axis as marked on the figure by $\hat{L}_z$.}
	\label{fig:OAMAxis}
\end{figure}

In \figref{fig:OAMAxis} the two laser fields considered in this work are depicted. The circular field propagates along the z or $\hat{e}_{||}$ direction, while the linear field propagates along the x direction. The angle relating to the final OAM of the photoelectron is chosen to be fixed for both fields around the z-axis, while its conjugate variable, the angle $\phi$, is marked on the figure. This convention will be used throughout the article. Note that, for any vector $\mathbf{v}$, the notation $v_{\perp}$ is used for the radial component in a cylindrical coordinate system, which is the projection of the vector onto the xy-plane, given by $v_{\perp}=\sqrt{v_x^2+v_y^2}$.

\subsection{Matrix element calculations for a general field }
\label{sec:matrix_element:general}
In order to proceed we will collect $\phi$ dependent and independent parts to analytically perform the integral in \eqref{eq:VortexTransition1}. The upper limit of the action contains $\phi$ independent terms that contribute only a phase and thus may be neglected as well as the following $\phi$ dependent term
\begin{equation}
p_{\perp} (\cos(\phi) \alpha_x(t) + \sin(\phi) \alpha_y(t)),
\end{equation}
where $\bm{\alpha}(t)=\int_{-\infty}^{t} \Ab(\tau) d \tau$. As $t\rightarrow \infty$ for a `well behaved' pulse or monochromatic field this term will vanish. 
For the case of $\sin^2$ pulses, which will be considered later, this holds for an $N$-cycle pulse, where $N$ is an integer greater than one. To simplify matters we will only consider such `well behaved' laser fields.
Thus, the upper limit of the action can be entirely neglected here. 

The lower limit of the action, which governs the dynamics, can be split into two parts:
\begin{align}
S(p_{\parallel},p_{\perp},\phi,t')&=
S_A(p_{\parallel},p_{\perp},t')+S_B(p_{\perp},\phi,t')\notag,
\intertext{%
  where
  }
S_A(p_{\parallel},p_{\perp},t')&=\notag\\
&\hspace{-2cm}\left(\ip +\frac{1}{2}p_{\parallel}^2+\frac{1}{2}p_{\perp}^2\right)t'
+p_{\parallel}\alpha_{\parallel}(t')
+\frac{1}{2}\int d t'A^2(t')
\intertext{%
  is independent of $\phi$, and
  }
S_B(p_{\perp},\phi, t')&= p_{\perp}
\left(\alpha_x(t')\cos(\phi)+\alpha_y(t')\sin(\phi)\right)\notag\\
&=  p_{\perp}\alpha_{\perp}(t')\sin(\phi-\nu(t'))\, ,
\end{align}
encodes the dependence on the momentum azimuth angle $\phi$.
This dependence is in terms of the angle
\begin{align}
    \nu(t') &= \mu(t') - \pi/2,
    \intertext{where}
    \mu(t')&=\arctan(\alpha_y(t')/\alpha_x(t')),
\end{align}
which is related to the rotation of the laser field and magnitude
\begin{align}
    \alpha_{\perp}(t')&=\sqrt{\alpha_x(t')^2+\alpha_y(t')^2}
\end{align}
of the field integral $\bm{\alpha}(t')$. 

Finally, in order to remove the $\phi$ dependence from the bound-state matrix element, $d(\pb,t')=\braket{\pb+\Ab(t')|V|\psi_{0}}$, we decompose it into its Fourier series
\begin{equation}
d(\pb,t')=\sum_{m}e^{i m \phi} V_{m}(p_{\parallel},p_{\perp},t').
\label{eq:Fourierdip}
\end{equation}
If the bound state is that of an atomic target, then the matrix element will have only a limited number of nonzero relevant Fourier terms, and these will be easy to compute numerically applying the fast Fourier transform (FFT). 
The transition amplitude from \eqref{eq:VortexTransition1} may be written as
\begin{align}\label{eq:GenTransitionAmp1}
M_{l}(p_{\parallel},p_{\perp},t)&=\notag\\
&\hspace{-20mm}
-i\sum_{m}\int_{-\infty}^{t}dt'
e^{i S_A(p_{\parallel},p_{\perp},,t')} V_{m}(p_{\parallel},p_{\perp},t')
I^{\phi}_l(p_{\perp},t'),
\end{align}
where $I^{\phi}_l(p_{\perp},t')$ contains the $\phi$ integral and corresponding terms and is given by
\begin{align}
I^{\phi}_l(p_{\perp},t')&=\frac{i^l}{2\pi}\int_{-\pi}^{\pi}d\phi'
e^{-i(l-m)\phi'}
e^{i p_{\perp}\alpha_{\perp}(t')\sin\left[\phi'-\nu(t')\right]}\notag\\
&=i^{l}e^{-i (l-m) \nu(t')}
J_{l-m}(p_{\perp}\alpha_{\perp}(t')).
\label{eq:GenralIphi}
\end{align}
Now the full transition amplitude can be written as
\begin{align}
M_{l}(p_{\parallel},p_{\perp})&=
\notag\\
&\hspace{-1.8cm}
i^{l-1}\sum_m
\int_{-\infty}^{\infty}\!\!\!dt'
e^{i S_{l-m}(p_{\parallel},p_{\perp},t')}
V_{m}(p_{\parallel},p_{\perp},t')
J_{l-m}(p_{\perp}\alpha_{\perp}(t')),
\label{eq:FinalGeneral}
\intertext{where}
S_{l}(p_{\parallel},p_{\perp},t')&=\left(\ip +\frac{1}{2}p^2_{\parallel}+\frac{1}{2}p^2_{\perp}\right)t'
-l \nu(t')
\notag\\
&
+p_{\parallel}\alpha_{\parallel}(t')
+\frac{1}{2}\int d t'A^2(t').
\label{eq:FullAction}
\end{align}
Note that the OAM $l$ only appears in the action coupled with $\nu(t')$, which mediates interaction of the laser field with the OAM of the electron. The time-varying `angle' $\nu(t')$ can be interpreted as the dynamical rotational action of the field on the OAM of the photoelectron. For a monochromatic field, we will see that $\nu(t')=\omega t$. We can rewrite $\nu(t')$ in terms of its derivative to gain more insight
\begin{equation}
    \nu'(t') = \frac{\left(\bm{\alpha}(t')\times\Ab(t')\right)\cdot \hat{e}_{\parallel}}
                    {\alpha_{\perp}(t')^2}.
\end{equation}
This makes the rotational nature and link to the spin of the field \cite{yang2020} more apparent.
We will now examine a few special cases that will significantly simplify the transition amplitude.

\subsection{Matrix element calculations for a linear field }
\label{sec:matrix_element:linear}
In this section we will consider a laser field, which only has components in the $\hat{e_{\parallel}}$ direction in the cylindrical coordinate system (i.e. perpendicular to the xy-plane see \figref{fig:OAMAxis} for more details).
Thus, the field has no $\phi$ dependence and only $S_A(p,\theta_p,t,t')$ contributes to the action. Furthermore, given that $\alpha_{\perp}(t')\rightarrow0$, the Bessel function will become a Kronecker delta
\begin{equation}
    J_{l-m}(p_{\perp}\alpha_{\perp}(t'))\rightarrow\delta_{lm},\notag
\end{equation}
and this can be demonstrated by re-evaluating \eqref{eq:GenralIphi}   
\begin{equation}\label{eq:linearIphi}
I^{\mathrm{linear}}_{\phi}=\frac{1}{2\pi}\int_{-\pi}^{\pi} e^{-i(m-l)\phi'}d\phi' 
=\delta_{l,m} \,.
\end{equation}
This is a selection rule, enforcing $l=m$, given by the symmetry of the problem.  Substituting this back into the transition amplitude leaves   
\begin{equation}\label{eq:LinTransitionAmplitude}
M_{l}(p_{\parallel},p_{\perp},)=i^{l-1}\int_{-\infty}^{\infty}dt'
e^{i S_A(p_{\parallel},p_{\perp}, t')} V_{l}(p_{\parallel},p_{\perp},t') \,.
\end{equation}
The selection rule comes from the exponential factor of the bound state and OAM of the free electron, which indicates that there will be a one-to-one correspondence between the two. Thus, if there is only one Fourier term $m=m_0$ in Eq.~(\ref{eq:Fourierdip}), the OAM of the photoelectron will be $l=m_0$.   
Aside from this conservation of angular momentum, the form of the OAM distribution is unchanged from the original SFA formalism for the plane wave momentum state.

\subsection{Matrix element calculations for a circular field}
\label{sec:matrix_element:circular}
In this section we turn to the OAM transition amplitude for a monochromatic circular field, where there is now a $\phi$ dependence in the semi-classical action.     
The form of vector potential for a circular monochromatic field is given by
\begin{equation}
\Ab(t)=-\sqrt{2 \up} 
\left(\cos(\omega t)\boldsymbol{e}_{x}
+\sin(\omega t)\boldsymbol{e}_{y}\right),
\label{eq:VecPotCirc}
\end{equation}
for more details on the orientation of the field see \figref{fig:OAMAxis}.
For this field in \eqref{eq:FinalGeneral} $\alpha_{\perp}(t') = \sqrt{2\up}/\omega$ and $\nu(t')=\omega t'$. Thus, the action from \eqref{eq:FullAction} becomes linear in $t'$ such that    
\begin{align}
S_{l-m}(p_{\parallel},p_{\perp},t')&=\chi_{l-m}(p_{\parallel},p_{\perp}) t',
\intertext{where}
\chi_{l-m}(p_{\parallel},p_{\perp})&=\ip+\up +\frac{1}{2}(p^2_{\parallel}+p^2_{\perp})
-(l-m)\omega.
\end{align}
The transition amplitude is then
\begin{align}\label{eq:CircTransitionAmp1}
M_{l}(p_{\parallel},p_{\perp})=&i^{l-1}\sum_m
J_{l-m}\left(p_{\perp}\sqrt{2\up}/\omega\right)\notag\\
&\times\int_{-\infty}^{\infty}dt'
e^{i \chi_{l-m}(p_{\parallel},p_{\perp}) t'} V_{m}(p_{\parallel},p_{\perp},t').
\end{align}
The integral acts to Fourier transform the prefactor term $\tilde{V}_{\pb0}$ to give
\begin{align}
M_{l}(p_{\parallel},p_{\perp})=i^{l-1}\sum_m
J_{l-m}\left(p_{\perp}\sqrt{2\up}/\omega\right)
\hat{V}_{m}(p_{\parallel},p_{\perp},\chi_{l-m})
\end{align}
where $\hat{V}_{m}(p_{\parallel},p_{\perp},\chi_{l-m})$ is the Fourier transform of $V_{m}(p_{\parallel},p_{\perp},t')$ with the frequency $\chi_{l-m}(p_{\parallel},p_{\perp})$.

To further simplify the problem, we will now consider the case with a very simple bound state, where the matrix elements have only  one Fourier series term for $m=0$, as would be the case for a simple $s$-state or the bound state for a zero range potential. In the example of a zero range potential \cite{Becker1994} the matrix element of the bound state is given simply by a constant dependent on the ionisation potential $V_0 = \sqrt{2\pi \sqrt{2 \ip}}$.
Now \eqref{eq:CircTransitionAmp1} becomes
\begin{align}
M_{l}(p_{\parallel},p_{\perp})=&2\pi V_0 i^{l-1}
J_{l-m}\left(p_{\perp}\sqrt{2\up}/\omega\right)\notag\\
&\times\delta\left(\chi_{l-m}(p_{\parallel},p_{\perp})\right).
\label{eq:CircTransitionAmp2}
\end{align}
Thus, for the case of monochromatic field with a simple bound state we arrive at another conservation equation
\begin{equation}
\ip+\up +\frac{1}{2}(p^2_{\parallel}+p^2_{\perp})
-(l-m)\omega=0.
\label{eq:CircularConservation}
\end{equation}
This is directly related to the semi-classical condition for ATI peaks \cite{freeman1987}, which can be interpreted forming due to additional photons absorbed by the photoelectron in the continuum, beyond that required for ionisation. In this case, however, each ATI peak will correspond to a different value of OAM, which will be shifted by the quantum magnetic number $m$.
This has interesting implications as it suggests that different values of the OAM  will be localised to specific energy regions in photoelectron emission spectra for ionisation via circularly polarised light. It is this idea that was exploited to produce interference vortices in our recent work \cite{Maxwell2020Faraday}. 

\subsection{Exploiting the saddle point approximation}
\label{sec:matrix_element:circularSPA}
In this subsection we discuss an alternative way to compute the OAM-SFA transition amplitude. In order to compute the OAM transition amplitude we employ the saddle point approximation for \eqref{eq:transition-ampl-plane} and then analytically perform the integral over $\phi$. Applying the saddle point approximation and ignoring the bound state prefactor \eqref{eq:transition-ampl-plane} becomes
\begin{align}
M(p_{\parallel},p_{\perp},\phi)&=
\sum_{s=0}^{N-1}\sqrt{\frac{2\pi i}
	{\partial^2 S/\partial t^2|_{t=t_s}}}
\exp \left[i S(p_{\parallel},p_{\perp},\phi,t_s) \right],
\label{eq:Transition_SPA}
\end{align}
where $N$ is the number of laser cycle considered and $t_s$ is given by
\begin{align}
\left(\pb+\Ab(t_s) \right)^2=-2\ip.
\label{eq:Times_SPA}
\end{align}
For a general field $t_s$ will depend on $\phi$ in a nontrivial way so the integral over $\phi$ can not be performed analytically. However, for a monochromatic circular field the dependence on $\phi$ is linear and we can write $t_s$ in terms of $t'_s$, which is independent of $\phi$
\begin{align}
\omega t_s(p_{\parallel},p_{\perp},\phi) &= \phi +\omega t'_s(p_{\parallel},p_{\perp}) &\quad (\mathrm{mod}\quad 2\pi).
\end{align}
Substituting this into the action leads to
\begin{align}
S(p_{\parallel}, p_{\perp}, \phi,t'_s)&=
\left(\ip+\up+\frac{1}{2}(p_{\parallel}^2 + p_{\perp}^2) \right)
\left( \frac{\phi}{\omega}+t'_s \right)\notag\\
&-\frac{\sqrt{2\up}}{\omega}p_{\perp}
\sin\left(\omega t'_s\right)\\
&=S(p_{\parallel},p_{\perp},0,t'_s)\notag \\
&+\left(\ip+\up+\frac{1}{2}(p_{\parallel}^2 + p_{\perp}^2) \right)\frac{\phi}{\omega}.
\end{align}
Thus, substituting the action and transition amplitude into \eqref{eq:FinalGeneral} leads to the expression
\begin{align}
M_{l}(p_{\parallel},p_{\perp})&=\sum_{s=0}^{N-1} \sqrt{\frac{2\pi i}
	{\partial^2 S/\partial {t'}_{s}^2}}
\exp(i S(p_{\parallel},p_{\perp},0,t'_s))\notag\\
&\hspace{-0.75cm}\times\frac{1}{2\pi}\int_{-\pi}^{\pi}d\phi' 
\exp\left[ i (\ip+\up+1/2 p^2)(\phi'/\omega)-i\phi' l \right]\\
&=\sum_{s=0}^{N-1} \sqrt{\frac{2\pi i}
	{\partial^2 S/\partial {t'}_{s}^2}}
\exp(i S(p_{\parallel},p_{\perp},0,t'_s))\notag\\
&\times
\mathrm{sinc}\left[(\ip+\up+1/2 p^2 -\omega l)(\pi/\omega) \right].
\end{align}
\begin{figure}
	\centering
	\includegraphics[width=\linewidth]{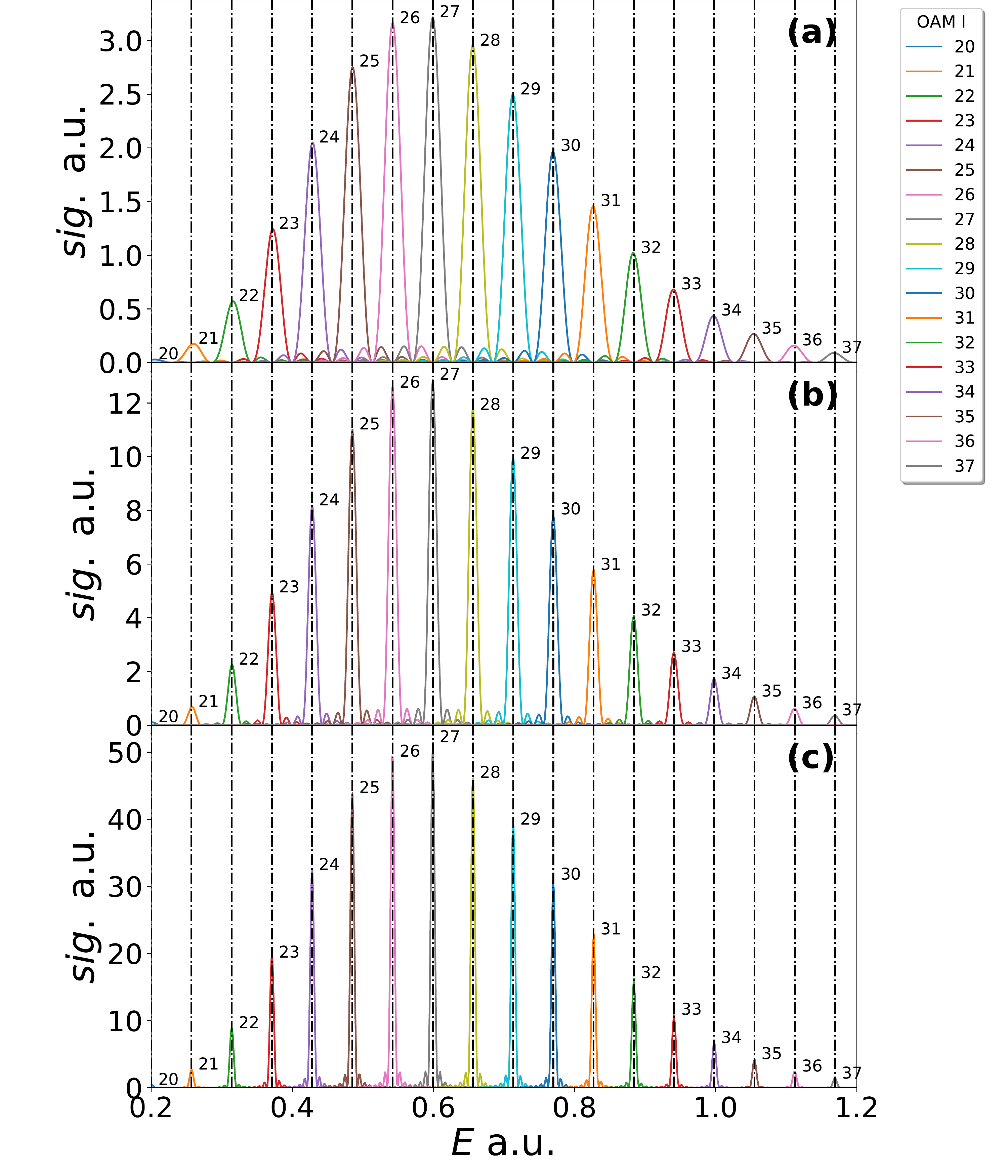}
	\caption{The result of \eqref{eq:OAM_SPA_Analytical} is plotted  for 2 [panel (a)], 4 [panel (b)] and 8 [panel (c)] ionisation events (given by N). 
		The laser field intensity employed is $I=2\times10^{14}$~W/cm$^2$ and a wavelength $\lambda=800$~nm, corresponding to $\up=0.44$~a.u. and $\omega=0.057$~a.u. for the ponderomotive energy and angular frequency, respectively.
		The distributions are plotted vs energy in order to easily see the condition (\ref{eq:CircularConservation}), which is marked by the vertical dashed lines on each figure and has constant spacing in energy.  The OAM l values 20 to 37 are included as shown in the figure legend. The bound state prefactor has been neglected in the calculation.
	}
	\label{fig:AnalyticPlot}
\end{figure}
Note, the prefactor in the square root can be determined to be independent of $\phi'$, thus it is outside of the $\phi'$ integral.
The sum leads to the contribution of one identical ionisation event per laser cycle, due to the field being monochromatic. Thus, the sum can be evaluated analytically to give the following
\begin{align}
M_{l}(p_{\parallel},p_{\perp})&=
\mathrm{\Omega}_{N}(p_{\parallel},p_{\perp}) 
\sqrt{\frac{2\pi i}{\partial^2 S/\partial {t'}_{s}^2}}
\exp(i S(p_{\parallel},p_{\perp},0,t'_0))\notag\\
&\times\mathrm{sinc}\left[\frac{\chi_l\pi}{\omega}\right] \,,
\label{eq:OAM_SPA_Analytical}
\intertext{where $t'_0$ denotes the solution in the first laser cycle and}
\mathrm{\Omega}_{N}(p_{\parallel},p_{\perp})&=
\sum_{n=0}^{N-1} \exp\left( 2\pi i n \chi_0/\omega \right)\notag\\
&=\frac{\exp\left[2\pi i N\chi_0/\omega\right]-1}
{\exp\left[2\pi i  \chi_0/\omega\right]-1}
\end{align}
as previously demonstrated in \cite{maxwell2017,arbo2010}.
In the limit $N\rightarrow \infty$ $\mathrm{\Omega}_{N}(p_{\parallel},p_{\perp})$ becomes a Dirac comb
\begin{align}
 \lim_{N\rightarrow\infty} \mathrm{\Omega}_{N}(p_{\parallel},p_{\perp})
 =\sum_{n=0}^{\infty} \delta\left(\chi_0- n\omega  \right).
\end{align}
The Dirac delta functions lead to the following replacement in the argument of the $\mathrm{sinc}$ function
\begin{align}
    \mathrm{sinc}\left[\frac{\chi_l\pi}{\omega}\right]
    &\rightarrow
    \mathrm{sinc}\left[\pi(n-l)\right]
    =\delta_{nl}.
\end{align}
Thus, this leads to 
\begin{equation}
  M_{l}(p_{\parallel},p_{\perp})=  
  \sqrt{\frac{2\pi i}{\partial^2 S/\partial {t'}_{s}^2}}
  \exp(i S(p_{\parallel},p_{\perp},0,t'_0))
  \delta\left(\chi_l(p_{\parallel},p_{\perp})\right),
\end{equation}
which gives the same condition as \eqref{eq:CircularConservation} but has different prefactors due to the saddle point approximation. Taking $N$ as a finite fixed value we are able to plot the result, which can be considered a rough approximation to employing a laser pulse of $N$ cycles, which has been done in \figref{fig:AnalyticPlot}.
The plot is over energy in order to make clear that each peak is separated with equal energy spacing, as dictated by \eqref{eq:CircularConservation}. As the number of laser cycles goes from 2 to 8 the OAM peaks get sharper, approaching the delta functions predicted by \eqref{eq:CircTransitionAmp2}. The bound state prefactor is neglected in \figref{fig:AnalyticPlot}, but it would be expected that including this should shift the peak by the value of the magnetic quantum number. This behaviour was in fact observed in \cite{Maxwell2020Faraday}. The analytical formalism we have derived is able to give the core properties of the OAM for strong field ionisation, however, in an actual experiment the laser will have a pulse envelope and thus a distribution of frequencies. In the next section, using numerical computations, we explore the effect this has on the OAM distribution.

\section{Numerical computations}
\label{sec:experiment_detection}
\begin{figure}
	\includegraphics[width=\columnwidth]{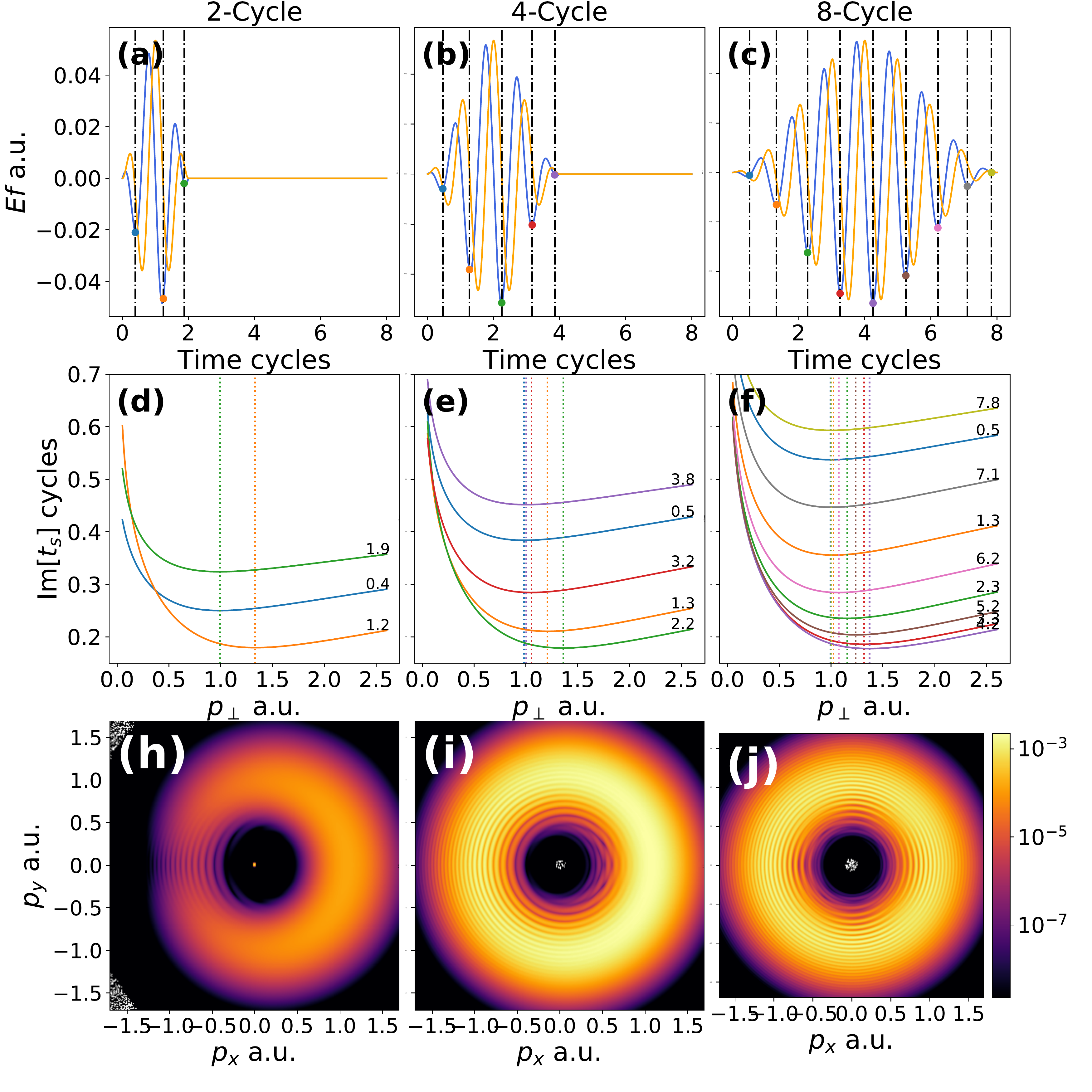}
	\caption{The laser pulse (first row), imaginary parts of ionisation times (second row) and momentum distributions (final row). In the first row the 2-cycle, 4-cycle and 8-cycle laser fields are plotted in panels (a), (b) and (c) respectively. The x [y] component of the laser field is plotted by the blue [orange] line. The real part of the times of ionisation are given by the vertical dashed lines and coloured spots are used to mark where this intersects with the x component of the laser field.
	These times are found by solving \eqref{eq:Times_SPA}.
	The imaginary parts of the solutions are plotted in the middle row, panels (d)--(f) vs the perpendicular momentum coordinate $p_\perp$, the colours of each line corresponds to the colour of the spots in the panels above, the real parts are also explicitly given (in units of laser cycles) on the right hand side of each panel. The minima of the imaginary part of the times of ionisation are marked by dotted lines. 
	The photoelectron momentum distributions are plotted along the bottom row, panels (h)--(i), for the momentum components in $xy$-plane and $p_z=0.1$~a.u., computed using \eqref{eq:Transition_SPA} with saddle point from \eqref{eq:Times_SPA} for each laser pulse considered.
	}
	\label{fig:MomentumDistribusions}
\end{figure}

In this section, we will examine the effect of a pulse envelope on the OAM distributions. We will employ 2-cycle, 4-cycle and 8-cycle laser pulses as in the previous section. Employing a $\sin^2$ envelope, we set the vector potential to be
\begin{equation}
\Ab(t)=-\sqrt{2\up}\sin\left(\frac{\omega t}{2 N}\right)^2 
\left(\cos(\omega t)\boldsymbol{e}_{x}
+\sin(\omega t)\boldsymbol{e}_{y}\right).
\end{equation}
The numerical computation of OAM distributions closely follows the methodology of the previous section, using the saddle point approximation to compute the \textit{plane-wave} transition amplitude (as in \eqref{eq:Transition_SPA}) and then \emph{numerically} computing the $\phi$ integral from \eqref{eq:FinalGeneral}. The $\phi$ integral can be computed very efficiently using the FFT algorithm.
The saddle point solutions are computed using \eqref{eq:Times_SPA}. The solutions can also be found very efficiently by transforming the equation to a polynomial of order $2N+2$ and finding the roots, as outlined in \cite{Jasarevic2020}. This means there will be $N+1$ valid solutions, see \figref{fig:MomentumDistribusions}.

\begin{figure}
	\includegraphics[width=\columnwidth]{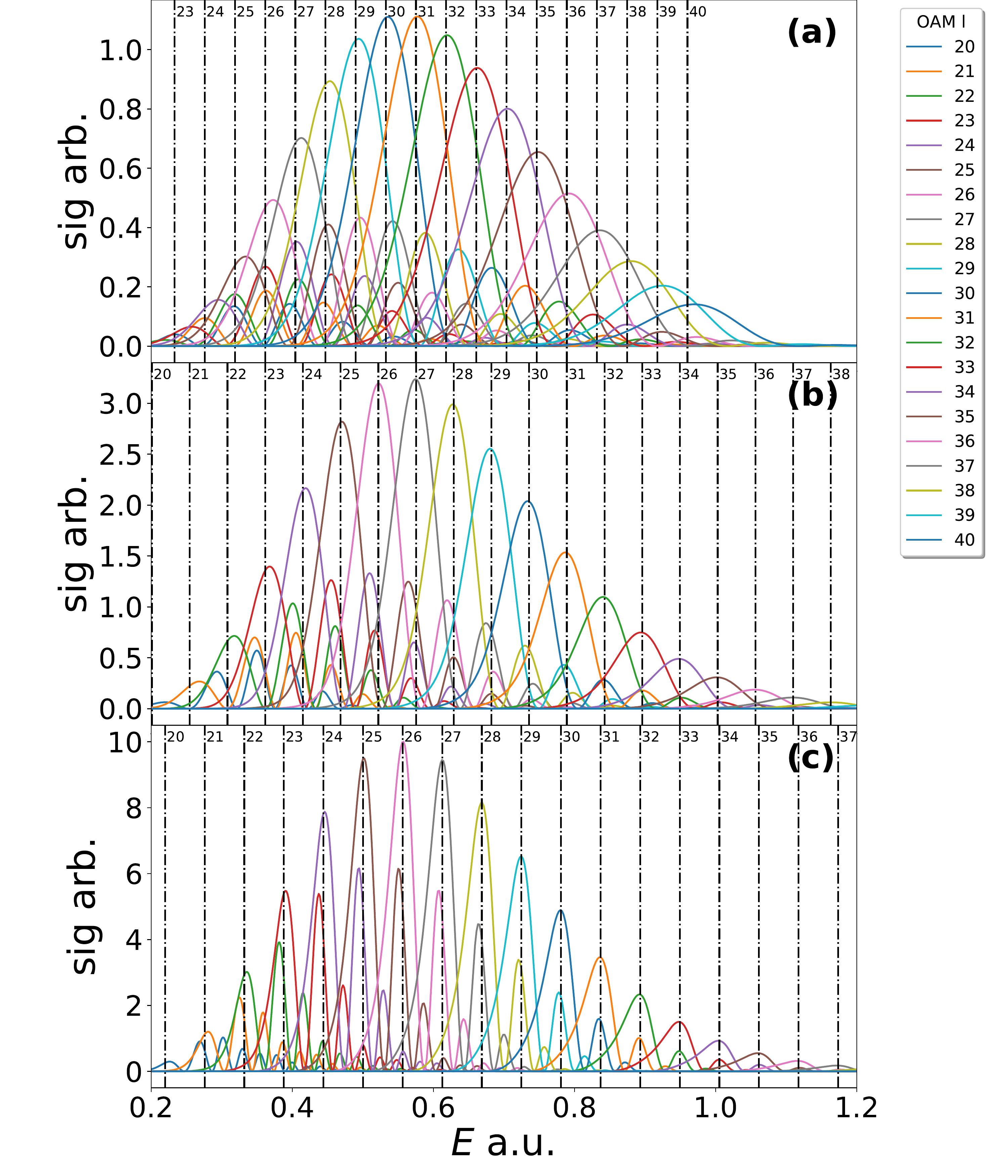}
	\caption{The OAM distributions are shown for a 2-, 4- and 8-cycle $\sin^2$ laser pulse, panels (a), (b) and (c), respectively. The peak intensity, wavelength and ionisation potential is the same as that used in \figref{fig:AnalyticPlot}. The bound state prefactors are neglected. The dot-dashed lines correspond to the condition \eqref{eq:CircularConservation} with the following effective frequencies used $3/4 \omega$, $15/16 \omega$ and $63/54 \omega$, for panels (a), (b) and (c), respectively. The condition has been slightly shifted in each case so that the central peak lines up with the corresponding $l$ value. The OAM values utilised are from $l=20$ to $l=40$.
	}
	\label{fig:OAMDistribusions}
\end{figure}

In \figref{fig:MomentumDistribusions} the 2-cycle, 4-cycle and 8-cycle results are shown. In the first column the laser field is plotted with the real parts of the times of ionisation indicated by the vertical dashed lines.  In the second column the imaginary parts of the ionisation times are plotted vs the radial momentum coordinate. The minima of these times (see the dotted line in \figref{fig:OAMDistribusions}) is a predictor for where the momentum distribution will be maximum. The momentum distributions are plotted in the final row for each laser pulse. As suggested by the imaginary part of the times, the peak of these distributions is away from the centre forming doughnut shapes. Interference between different paths can be seen as faint circular fringes. 
The following references \cite{Shearer2011, Jasarevic2020,Barth2013} can give some further insight into these solutions and the method.
Such distributions bear some similarity with the attoclock \cite{Eckle2008,Eckle2008Streaking,Pfeiffer2011,Pfeiffer2012,Landsman2014,Torlina2015,Sainadh2019,SatyaSainadh2020}, where an elliptical nearly circular field is used to relate the electron emission angle to the tunnelling time. In this work we are relating the photoelectron OAM to the emission energy, which are the conjugate variables of angle and time, respectively.

In \figref{fig:OAMDistribusions} the OAM distributions for the three laser pulses used in \figref{fig:MomentumDistribusions} are shown. The distribution of frequencies in the $\sin^2$ pulse means that we should not expect \eqref{eq:CircularConservation} to hold. In panel (a), the OAM distribution is plotted for a 2-cycle pulse.  There are still specific OAM peaks in each energy region, however now they significantly overlap. Furthermore, the central peak at around $0.6$~a.u. corresponds to $l=31$ instead of $l=27$ as in the monochromatic case. This is because the spacing between the peaks is reduced. The dot-dashed lines in \figref{fig:OAMDistribusions} panel (a) use a spacing of $\frac{3}{4}\omega$. This closely matches most of the peaks, but for increasing $l$ above $l=31$, the spacing drifts to higher values, while for decreasing $l$ below that of $l=31$ the spacing drifts to lower values. For the longer pulses, 4 and 8 cycles, the distributions move closer to the monochromatic case, with the $l=27$ peak moving towards its previous position of $0.6$~a.u. and the spacing between peaks getting closer to $\omega$. However, the spacing is still below this, with spacing of $\frac{15}{16}\omega$ and $\frac{63}{64}\omega$ being used for the dot-dashed lines for 4 and 8 cycles, respectively. Note that in all cases as well as altering the frequency in condition \eqref{eq:CircularConservation} a small shift was required to align the dot dashed lines to the correct central peak  (reducing for longer pulses), see the caption of \figref{fig:OAMDistribusions} for more details. 

A clear pattern is visible in the spacing of the OAM distribution of 2-, 4- and 8-cycle pulse, and it is possible to analytically derive this spacing dependence. In the monochromatic case, the spacing between OAM peaks is given by $\nu(t')=\omega t'$, which leads to an $\omega$ spacing in energy in the condition \eqref{eq:CircularConservation}. There is not such a simple relation for $\nu(t')$ when employing a $\sin^2$ pulse, but through a Taylor expansion we can derive such an expression. A first order Taylor expansion about the peak of the pulse ($t=\pi N/\omega)$ gives $\nu(t')=\omega^*_{N} t'$ in terms of an effective frequency $\omega^*_{N}$, dependent on the number of laser cycles. Note a constant term has been discarded as this only contributes an overall phase and does not affect the OAM peak separation. The effective frequency can be calculated to be
\begin{align}
\omega^*_{N} = \nu'(\pi N/\omega)=
\begin{cases}
\left( 1-\frac{1}{N^2} \right)\omega, & \text{if } N \text{ is even}\\
\omega & \text{otherwise}
\end{cases}\,,
\end{align}
which matches the peak spacing seen in 2-cycle, 4-cycle and 8-cycle pulse in \figref{fig:OAMDistribusions}.

\begin{figure}
	\includegraphics[width=\linewidth]{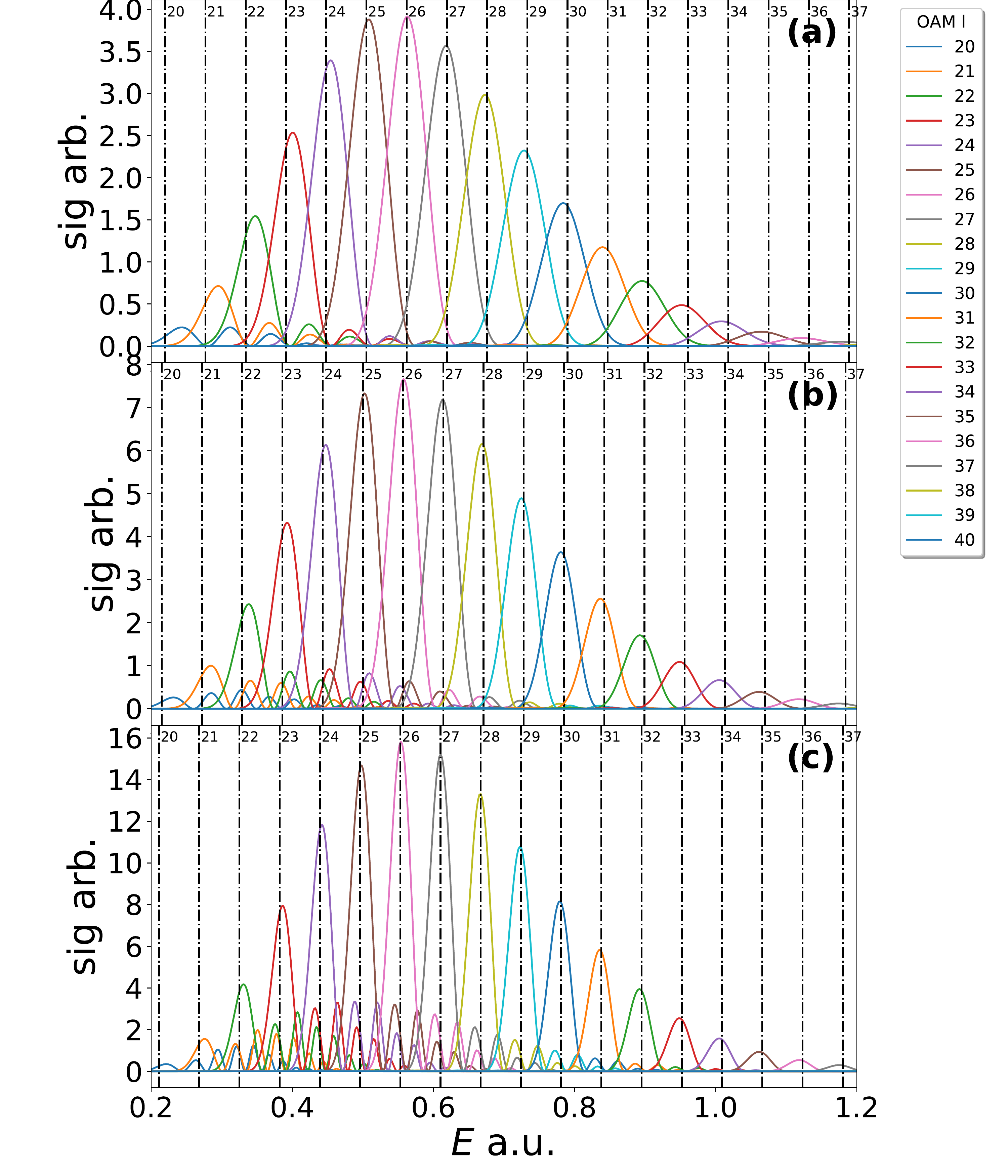}
	\caption{The same as \figref{fig:OAMDistribusions} except 3-, 5- and 9-cycle $\sin^2$ laser pulses have been employed. The dot-dashed lines use the carrier frequency for the spacing and are again slightly shifted so that \eqref{eq:CircularConservation} to match the central peak.}
	\label{fig:OAMDistribusions_Odd}
\end{figure}
It is interesting to note that, if the number of cycles $N$ is odd, the peaks remain separated at the carrier frequency. This is exemplified in \figref{fig:OAMDistribusions_Odd}, which shows the OAM distributions for 3-cycle, 5-cycle and 9-cycle laser pulses. The dot-dashed lines in this figure are all separated by the carrier frequency and the corresponding peaks in the distributions line up with this well. 
It is possible to see some drift away from the central peak in panel (a), as the spacing is slightly smaller than the carrier frequency. These shifts are simply because the condition, although accurate, it is no longer exact and thus there is a increasing drift for higher $l$ values.

Beyond the spacing between peaks there are further differences between \figref{fig:OAMDistribusions} and \figref{fig:OAMDistribusions_Odd}. 
Namely, in \figref{fig:OAMDistribusions} considerable secondary peaks can be observed, which for in cases where the OAM peaks are low [see $l=23$ in panel (c)] the secondary peaks can be nearly as high as the primary peak [or even higher in extreme cases e.g. $l=22$ in panel (c)]. However, the secondary peaks always occur in the regions where there is a more dominant OAM peak corresponding to another $l$ value. In contrast, for an odd number of cycles, \figref{fig:OAMDistribusions_Odd}, the secondary peaks are considerably lower, not playing much of a role at all. 
These features suggest that, if a clean well separated (in energy) OAM distribution is required, it would be much better to utilise an odd number of laser cycles in the laser pulse envelope.
Furthermore, in the even-cycle case the peaks are slightly asymmetric, with a longer tail on one side than the other, while for an odd number of cycles the peaks are symmetric. These differences suggest one may be able use the OAM of the photoelectrons to detect if the laser field had a closer to odd or even number of cycles. 
However, additional effects such as the intensity variation over the laser focal volume (see Ref.~\cite{Kopold2002} for an example) would need to be considered first and, as previously stated, the measurement of the OAM of photoelectrons in strong field experiment is an open problem \cite{Maxwell2020Faraday,Grillo2017}.

\section{Conclusions}
\label{sec:conclusions}
In this study we have developed a new version of the strong field approximation (SFA) to explore the orbital angular momentum of photoelectrons undergoing strong field ionisation. Employing this model we are able to derive analytic conditions relating to selection rules/ conservation laws about the system. In the case of a linear field we demonstrate that the photoelectron OAM is determined by the magnetic quantum number of the initial bound state, while for a circular field we show that a range of OAMs are possible, which will occur in well-defined energy regions. We derive an analytic condition that relates their position to the ATI peaks. Computations using a $sin^2$ laser pulse demonstrate that this condition continues to be accurate even for very short pulses but the separation of the peaks is now described by an analytically derived effective frequency as opposed to the carrier frequency of the laser field. We find that employing an odd or even number of laser cycles has a marked effect on the OAM distributions, leading to two classes of effective frequency for even and odd pulses.  

The present work is an interesting example of how the ellipticity and time profile of the field add dynamic shifts $\nu(t')$ to angular variables intrinsic to the target, such as the OAM of the electron.
This effect bears some similarity with previous work on the angular properties of the photoelectrons. In one example, the tunnelling angle, derived in \cite{barth2011}, was used to show the preferential tunnel ionisation of `counter-rotating' electrons with circularly polarised fields. An electron's angle of return also leaves imprints in HHG spectra \cite{Kitzler2008,Das2013}. These angular shifts were incorporated in a purely structural two-centre interference condition for HHG in diatomic molecules  \cite{Das2013}, and could be made visible by exploiting macroscopic propagation \cite{Das2015}.   A key difference is that these articles computed angles related to the velocity of the electron, whereas in the present work $\nu(t')$ is related only to the rotation of the field and represents the interaction of the laser field with the OAM of the electron. It could be said to be the `stirring' action of the field upon the electron at the moment of ionisation.

The method proposed here is general and could be extended to initial states with arbitrary angular momentum and other types of tailored fields.
However, an open question is the role of the residual binding potential. In \cite{Maxwell2020Faraday} good agreement was found between the SFA and the TDSE solvers Qprop and RMT but the longer wavelengths used in this study may lead to more significant Coulomb effects.
This question has been addressed in the attoclock setup, which (as previously stated) deals with the conjugate variables of the emission angle and corresponding ionisation time.
In this setting the Coulomb potential has been shown to shift the photoelectron emission angle (see e.g. \cite{Torlina2015}), which may have a profound effect on the interpretation of the tunnelling time (see \cite{Hofmann2019,SatyaSainadh2020} for reviews on the attoclock). Thus, it should be expected that the Coulomb potential will also shift the OAM of the photoelectron.  This may be addressed by incorporating the OAM into SFA-like models which account for the binding potential, such as the Coulomb quantum-orbit strong field approximation \cite{lai2015,maxwell2017,faria_it_2020} as well as performing comparison with TDSE solvers.

The OAM decomposition of the final photoelectron wavepacket presented here reveals the angular momentum content along a chosen direction, associated with the OAM that may be carried by freely propagating particles. The electron wavepacket does not form a beam. Therefore, the wavepacket cannot be envisaged as a single vortex beam propagating along the OAM axis, but rather a superposition of many different Bessel beam vortex states with different OAMs. This can be exploited either in strong field measurement \cite{Maxwell2020Faraday,Grillo2017}, where such distributions over OAM may reveal unique properties about the target system or alternatively, could be used to produce a custom electron vortex beam \cite{lloyd2017}. Such as scheme could exploit the relationship between the ATI peaks and the OAM to filter photoelectrons from a specific ATI peak in order to select a single OAM.

Another very important question is:  How about other types of fields, for more complex vortices? The response of matter to structured laser fields is a central subject of intense laser-matter physics, and is inevitably related to the OAM of matter.  
Recently, a novel pump-probe scheme incorporating OAM was theoretically demonstrated using a vortex IR beam and XUV pulse allowing for time-resolved photoionisation \cite{Giri2020}, while DeNinno \cite{DeNinno2020} demonstrated experimentally that a free electron matter wave, produced by an XUV pulse, is sensitive to a vortex IR beam. Similarly, angle-resolved attosecond streaking of twisted attosecond pulses has been recently proposed \cite{Ansari2020}.
In a condensed matter context, it was shown recently that THz laser pulses with circular polarisation induce transient Chern insulator in graphene \cite{McIver2020}. Similarly, linearly polarised pulses with OAM induce non-uniform Chern insulators \cite{Bhattacharya2020arXiv}. All these examples clearly illustrates that we are entering an era of OAM and more complex textures in laser-matter physics.

\section*{Acknowledgements}
The data and plotting scripts for all figures used in this manuscript is freely available at \href{https://doi.org/10.5281/zenodo.4551606}{Zenodo}. The code written for this manuscript has been made open source and is available as a release on \href{https://doi.org/10.5281/zenodo.5076167}{Zenodo} and on \href{https://github.com/asmaxwell/SFA_OAM_Circular}{GitHub}.

ASM acknowledges grant EP/P510270/1 and CFMF grant No.\ EP/J019143/1, both funded by the UK Engineering and Physical Sciences Research Council (EPSRC). ASM, EP, MC and ML acknowledge support from ERC AdG NOQIA, Agencia Estatal de Investigación (``Severo Ochoa'' Center of Excellence CEX\allowbreak{}2019-000910-S, Plan National FIDEUA PID2019-106901\allowbreak{}GB-I00/10.13039 / 501100011\allowbreak{}033, FPI), Fundació Privada Cellex, Fundació Mir-Puig, and from Generalitat de Catalunya (AGAUR Grant No.\ 2017 SGR 1341, CERCA program, QuantumCAT \_U16-011424, co-funded by ERDF Operational Program of Catalonia 2014-2020), MINECO-EU QUANTERA MAQS (funded by State Research Agency (AEI) PCI2019-111828-2 / 10.\allowbreak{}13039/\allowbreak{}501100011033), EU Horizon 2020 FET-OPEN OPTOLogic (Grant No 899794), and the National Science Centre, Poland-Symfonia Grant No.\ 2016/20/W/ST4/00314, Marie Skłodowska-Curie grant STRETCH No.\ 101029393, ``La Caixa'' Junior Leaders fellowships (ID100010434),  and EU Horizon 2020 under Marie Skłodowska-Curie grant agreement No 847648 (LCF/BQ/PI19/11690013, LCF/BQ/PI20/11760031,  LCF/BQ/PR20/11770012).

%
 \bibliographystyle{arthur}
 \interlinepenalty=10000
 \bibliography{OAM_SF_sorted.bib}
%
%
%

\end{document}